\documentclass[sigconf]{acmart}
\AtBeginDocument{%
  }

\copyrightyear{2025}
\acmYear{2025}
\setcopyright{cc}
\setcctype{by}

\usepackage{subcaption}
\usepackage{multirow}
\usepackage{textcomp}
\usepackage{float}

\def \cbox {$\bigcirc$}

\begin{document}

\title{Exploring Bichronous Collaboration in Virtual Environments}

\author{Alexander Giovannelli}
\email{agiovannelli@vt.edu}
\orcid{0000-0002-0265-8143}
\affiliation{%
  \institution{Virginia Tech}
  \state{VA}
  \country{USA}
}

\author{Shakiba Davari}
\email{sdn6@gatech.edu}
\orcid{0000-0003-3128-1979}
\affiliation{%
  \institution{Georgia Institute of Technology}
  \state{GA}
  \country{USA}
}

\author{Cherelle Connor}
\email{cconnor22@vt.edu}
\orcid{0000-0002-8072-4206}
\affiliation{%
  \institution{Virginia Tech}
  \state{VA}
  \country{USA}
}

\author{Fionn Murphy}
\email{fionncmurphy@vt.edu}
\orcid{0009-0002-2630-5366}
\affiliation{%
  \institution{Virginia Tech}
  \state{VA}
  \country{USA}
}

\author{Trey Davis}
\email{dtrey05@vt.edu}
\orcid{0009-0001-8291-8354}
\affiliation{%
  \institution{Virginia Tech}
  \state{VA}
  \country{USA}
}

\author{Haichao Miao}
\email{miao1@llnl.gov}
\orcid{0000-0001-6580-2918}
\affiliation{%
 \institution{LLNL}
 \state{CA}
 \country{USA}
}

\author{Vuthea Chheang}
\email{vuthea.chheang@sjsu.edu}
\orcid{0000-0001-5999-4968}
\affiliation{%
 \institution{San José State University}
 \state{CA}
 \country{USA}
}

\author{Brian Giera}
\email{giera1@llnl.gov}
\orcid{0000-0001-6543-7498}
\affiliation{%
 \institution{LLNL}
 \state{CA}
 \country{USA}
}

\author{Peer-Timo Bremer}
\email{bremer5@llnl.gov}
\orcid{0000-0003-4107-3831}
\affiliation{%
 \institution{LLNL}
 \state{CA}
 \country{USA}
}

\author{Doug A. Bowman}
\email{dbowman@vt.edu}
\orcid{0000-0003-0491-5067}
\affiliation{%
  \institution{Virginia Tech}
  \state{VA}
  \country{USA}
}

\renewcommand{\shortauthors}{Giovannelli et al.}

\begin{abstract}
Virtual environments (VEs) empower geographically distributed teams to collaborate on a shared project regardless of time.
Existing research has separately investigated collaborations within these VEs at the same time (i.e., synchronous) or different times (i.e., asynchronous).
In this work, we highlight the often-overlooked concept of bichronous collaboration and define it as the seamless integration of archived information during a real-time collaborative session.
We revisit the time-space matrix of computer-supported cooperative work (CSCW) and reclassify the time dimension as a continuum.
We describe a system that empowers collaboration across the temporal states of the time continuum within a VE during remote work.
We conducted a user study using the system to discover how the bichronous temporal state impacts the user experience during a collaborative inspection.
Findings indicate that the bichronous temporal state is beneficial to collaborative activities for information processing, but has drawbacks such as changed interaction and positioning behaviors in the VE.
\end{abstract}


\begin{teaserfigure}
  \includegraphics[width=.9\textwidth]{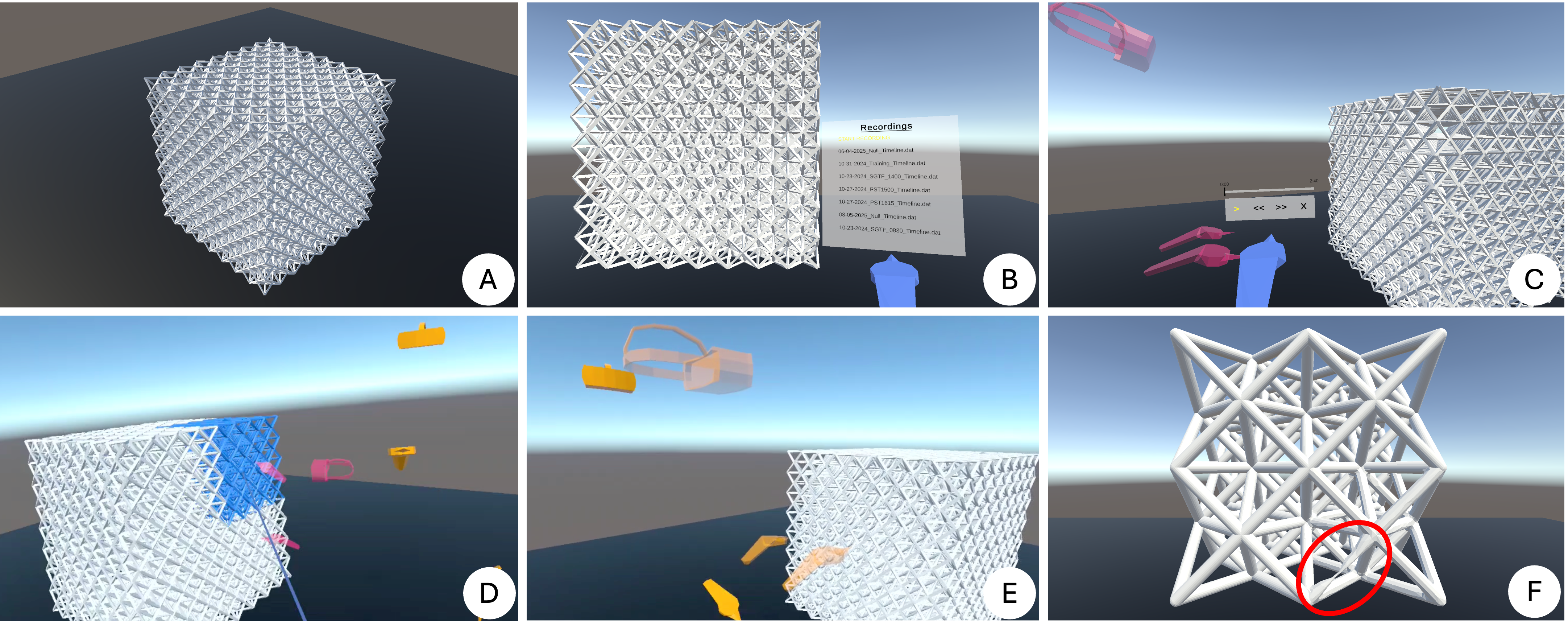}
  \centering
  \caption{View of the collaborative virtual environment from the desktop (A\&F) and the participant's field of view (B-E). A: The cuboid-lattice inspection object. B: Recorded avatar annotation menu. C: Asynchronous collaboration with playback control interface and pink, semi-transparent avatar annotation. D: Bichronous collaboration with pink, semi-transparent avatar annotation and orange, opaque live collaborator. E: Bichronous collaboration with live collaborator and their avatar annotation. F: An example defective strand in a cell of the cuboid-lattice inspection object.}
  \Description{}
  \label{fig:teaser}
\end{teaserfigure}

\received{14 July 2025}
\received[revised]{8 September 2025}
\received[accepted]{12 September 2025}

\maketitle

\section{Introduction}
As globalization continues to revolutionize how people work together, the ability to communicate effectively and efficiently between team members and stakeholders has grown in importance \cite{Morrison_VirtualTeamsLitReview2020}.
This has led to the common practice of using telecommunications and messaging platforms such as Zoom \cite{Zoom} and Slack \cite{Slack} to empower geographically dispersed teams to share information and coordinate future work at the same time (i.e., synchronously) and different times (i.e., asynchronously).
However, these platforms are still limited compared to the practice of working together in a shared physical space, where team members can communicate face-to-face while working with objects or documents \cite{Sellen_RemoteConversations1995}.

Extended Reality (XR) technologies offer a means to resolve previously identified issues such as collaborator representation \cite{Giovannelli_GesturesVsEmojis_2023}, content visualization \cite{Li_ARCritique2022}, and engagement \cite{Giovannelli_GuidedTours2025} prevalent in 2D telecommunications platforms.
By embodying users in a Collaborative Virtual Environment (CVE), collaborators can view and interact naturally with each other and complex visualizations \cite{Bowman_IRVE2003, Mayer_AsyncManualWork2022}.
For these reasons, commercial organizations have developed such CVEs for professional work (e.g., Spatial.io \cite{Spatial}) and social activities (e.g., VRChat \cite{VRChat}).

Due to the potential of CVEs to enable organizational work regardless of physical location, researchers have contributed significantly to understanding their nuances and capabilities in recent years \cite{Ens_RevisitingCollabMR2019}.
However, most existing contributions adhere strictly to Johansen's time-space matrix, which represents both time and space as binary dimensions: synchronous or asynchronous for time and co-located or remote for space \cite{Johansen_CollaborativeSystems1989}.
While the binary representation of the space dimension has been revisited and evolved into a continuum \cite{Thomas_PICC2023}, the time dimension has remained the same since its induction.
We argue that the time dimension must also be revisited, as time during collaboration is a continuum with extremes of asynchronous and synchronous work.
Although these extreme temporal states generally characterized CSCW in its early years \cite{Neale_CSCWModels2004}, recent work suggests that in the continuum, we should consider \textit{bichronous collaboration}, a mixed temporal state with aspects of both synchronous and asynchronous collaboration. 
We define bichronous collaboration as the seamless integration of archived information during a real-time collaborative session.

In this paper, we describe our early work to investigate the bichronous temporal state in the context of a CVE.
We contribute the following: (i) the formal identification and definition of the collaborative time continuum; (ii) a prototype developed to investigate collaboration in Virtual Reality (VR) across the synchronous, asynchronous, and bichronous temporal states; (iii) findings from an evaluation conducted with the prototype detailing the advantages and limitations of the bichronous temporal state in the context of VR collaboration.

\section{Related Work}
CSCW has developed a substantial body of research on collaborative applications.
Although many manuscripts exist in this area of research, we focus our literature review on publications concerning XR collaboration systems.

\subsection{Synchronous Collaboration}
When performing task work during a synchronous collaboration in XR, existing research has examined how to direct attention towards active tasks \cite{Churchill_CollabVEs1998}.
One such method is by controlling the point of view of specific team members \cite{Valin_ViewShareCVE2001}.
Lanir et al. examined the impact of point of view control during an assistance scenario between a remote expert helper and local worker performing real-time manual construction tasks \cite{Lanir_PoVControl2013}.
They found that the point of view should generally be controlled by the more knowledgeable collaborator during the task.
Expanding on the importance of point of view during synchronous collaboration, Kim et al. and Teo et al. explored visual cue methods to direct attention to task actions using AR and 360\textdegree\ video views \cite{Kim_ARViewShare2013, Teo_MRVisualCues2019}.
Kim et al. found that drawn cues were more effective at conveying where to direct attention in a shared view than simple dot reticles for 2D viewports offered by the AR 2D-viewport, whereas Teo et al. indicated further work was needed to examine the influence of such cues in 360\textdegree\ video.

Aside from the live view of individual members, the representation of collaborators via avatars is integral for real-time collaboration using XR technologies \cite{Schafer_SyncSurveyXR2022}, as they influence communication and interaction within CVEs \cite{Heidicker_AvatarAppearance2017, Lindgren_Principles2023}.
Yoon et al. investigated the impact of body part visualization and realistic styling of avatars during collaboration on social presence and perception \cite{Yoon_EffectAvatarAppearance2019}. 
They found that styling did not impact sense of social presence and that full- or half-body visualizations were sufficient when observing collaborators during task work.
Pakanen et al. explored user preferences of body part visualization and styling when embodying rather than observing an avatar \cite{Pakanen_AvatarTelexistence2022}.
They found that participants generally preferred embodying photorealistic, full-body avatars due to their human-like representation and affordances for interaction.
These contributions indicate that task activities in immersive experiences must consider methods to direct collaborators' attention in the environment and avatar representations of collaborators to ensure effective synchronous collaboration.

\subsection{Asynchronous Collaboration}
When conducting asynchronous collaboration using XR technologies, it is critical for collaborators to be aware of what tasks have been done \cite{Tam_AsyncChangeAwareness2006}.
Annotations, defined as ``generic units of information ... that can consist of text, images, animations, sounds, and other forms'' \cite{Verlinden_VirtualAnnotation1993}, have been researched for their ability to spatially index and promote awareness of changes in XR systems for this reason \cite{Harmon_AnnotationSys1996, Irlitti_ChallengesAsync2016}. 
To make collaborators aware of asynchronous contributions, Marques et al. investigated various notification methods for remote experts to guide onsite technicians to spatial annotations \cite{Marques_AsyncNotifications2022}.
They found tactile cues provided greater worker attention and awareness to remote expert annotation updates.
Additionally, they later examined concerns of live technician annotation consumption ordering and step-by-step instruction generation for experts operating at different times, receiving positive feedback for their system from both types of users \cite{Marques_RemoteAsyncCollab2021}.

Chow et al. used avatar annotations (i.e., recordings of user avatar(s) in the VE capable of being played back by observing user(s)) in an effort to improve understanding of asynchronous contributions \cite{Chow_ChallengesAsyncCollabVR2019}.
They identified design considerations for these annotations, including accommodating viewpoint control, supporting navigation of their playback, and coordinating proxemic behaviors.
Wang et al. developed a similar communication tool for asynchronous collaboration for ease of environment referencing in interior design \cite{Wang_VRReplay2019}.
From a resulting study with their tool, users preferred the use of the avatar recordings for communication clarity, perception of partners, and perceived performance.
These contributions demonstrate the importance of creation and retention of information via annotations for time-distributed observers to become aware of and understand task activities in asynchronous collaboration.

\subsection{Bichronous Collaboration}
Bichronous collaboration is a term that has only recently been coined in the field of e-learning research as ``the blending of both asynchronous and synchronous online learning, where students can participate in anytime, anywhere learning during the asynchronous parts of the course but then participate in real-time activities for the synchronous sessions'' \cite{Florence_BichronousLearningDef2020, Brzezinska_BichronousLearning2022}.
However, to our knowledge, no work has explicitly investigated bichronous collaboration in the context of XR technologies.
While some XR systems have incorporated features that could support bichronous collaboration, they have not framed or analyzed their work through this lens.
The use cases for such systems have included reliving personal experiences with a partner \cite{Wang_AgainTogether2020}, conducting comprehensive literature reviews \cite{Tahmid_ColtCollab2023}, and joint viewing of avatar annotations for social behavior analysis \cite{Javerliat_PlumeRecordReplay2024}.\\

\noindent Bichronous collaboration has the potential to promote real-time discussion from collaborators inhabiting a given CVE by blending the attention directing properties of live peers via synchronous collaboration with the retention and understanding of expert insights via asynchronous collaboration.
For this reason, we developed our own VR system to investigate bichronous collaboration and explore its benefits and drawbacks.

\section{Rethinking the Collaborative Time Dichotomy} \label{CollabStates}
With the adoption of technology in collaborative activities, Johansen proposed a widely adopted matrix that defined a discrete, binary classification for collaboration across time and space \cite{Johansen_CollaborativeSystems1989}.
Space, referred to as ``place'' in the matrix, was described as where users conducted collaborative actions, being either at the same physical location (i.e., collocated) or by using remote media (e.g., teleconferencing).
Time was described as when users conducted collaborative actions, working together at the same time (i.e., synchronous) or at different times (i.e., asynchronous).

\begin{figure}
    \centering
    \includegraphics[width=\linewidth]{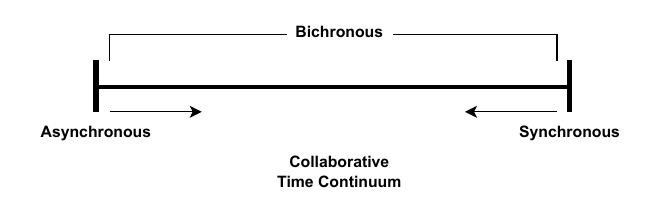}
    \caption{The proposed collaborative time continuum. The arrows represent the extremes of the continuum turning into the continuous range of the bichronous intermediary state.}
    \Description{}
    \label{fig:transitions}
\end{figure}

Many research contributions have been made to understand and support the four discrete combinations defined by the time-space matrix in the context of XR \cite{Ens_RevisitingCollabMR2019, Pidel_CollabSystematic2020, Ghamandi_CollaborationTaxonomy2023, Assaf_SurveyCollaboration2024}.
From these contributions, space has grown from its binary classifications to a continuum, spanning an intermediary state of ``hybrid locality'' in which there are both co-located and remote collaborators in the same collaborative session \cite{Benford_MRSharedSpaceBoundaries1998, Thomas_PICC2023}.
However, research about the time dimension in XR collaboration has only investigated its discrete states.
Johansen acknowledged that the time-space matrix was ``simple’’ and ``would evolve in a similar fashion’’ as technologies supporting collaborative work did before it \cite{Johansen_CollaborativeSystems1989}.

Synchronous collaboration occurs when two or more \textit{live collaborators} work with one another toward a common goal in real time.
Asynchronous collaboration occurs when authoring collaborator(s) create and store information that is reviewed at a different time by observing collaborator(s) via \textit{annotations}.
Such annotations allow members of the collaborating group to discover, engage with, and learn the progress of a task.
These temporal states of collaboration are distinguished by their reliance on either \textit{live collaborators} or \textit{annotations}.

\begin{figure*}
    \centering
    \includegraphics[width=.9\linewidth]{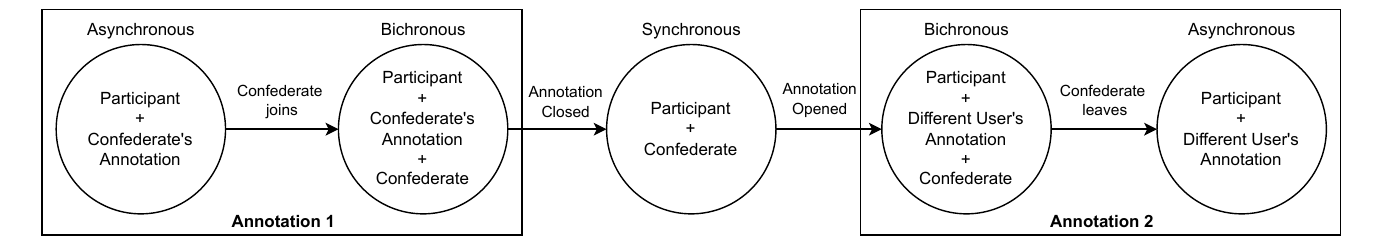}
    \caption{Visual representation of the study review task. Temporal states are fixed above the entities involved. Transitions and their causes are transcribed between the temporal states.}
    \Description{}
    \label{fig:task}
\end{figure*}

Bichronous collaboration is a blend of synchronous and asynchronous states, where information captured in \textit{annotations} is used during a real-time meeting consisting of two or more \textit{live collaborators}.
It provides ease of access to archived information such that many people are able to consume that information and discuss it in real time.
Thus, we argue that bichronous collaboration exists on a continuum between the asynchronous and synchronous extremes in the collaborative time continuum (Fig. \ref{fig:transitions}).
The extent to which the bichronous temporal state gravitates more towards synchronous or asynchronous extremes depends on the quantity of the annotation content inhabiting the scene during a live collaboration compared to the quantity of live collaborators.

Let us consider a set of transitions between temporal states in a scenario to showcase the proposed time continuum.
A work group, consisting of three workers (i.e., W1, W2, and W3), uses a CVE for quality assurance reviews (QARs) of products made by their organization.
W1 enters the CVE to perform a QAR on a product model, entering the asynchronous temporal state once they start to create and store their findings using annotations.
W1 then leaves the CVE.
Later in the day, W2 enters the CVE to continue the QAR.
W2 opens an annotation from W1 to discover what has been done and to determine what to inspect in more depth, thus entering the asynchronous temporal state.
W3 then enters the CVE while W2 is reviewing W1's annotation.
W3 joins W2 in the review and they discuss information provided by W1's annotation, thus entering the bichronous temporal state.
Since there is only one pair of live collaborators and one annotation, we consider this to be in the middle of the continuum between synchronous and asynchronous.
Continuing the example, not enough context is provided by W1's annotation, so another annotation from W1 is opened by W2 and W3.
This is still a bichronous collaborative temporal state, but because of the additional annotation, we consider this to be closer to the asynchronous extreme of the continuum.
After discussion regarding the second annotation's information, W2 and W3 close all of W1's annotations and resume the QAR, thus entering the synchronous temporal state.

\section{Experiment}
We conducted an exploratory study investigating the influence of temporal states during a collaborative scenario using a custom-developed bichronous collaboration prototype.
The prototype's use case focuses on additive manufacturing (AM), a recently identified specialty for engineering-based VR collaboration \cite{Chheang_AMInspect2024}.
In AM, it is common for workers to be distributed geographically across manufacturing and design facilities, requiring collaboration to take place remotely and across temporal states. 
We focused on the task of inspecting an AM part for defects, which often requires collaboration between design and manufacturing specialists.
The prototype allowed users to record their work in the VE as avatar annotations, and these recordings could be played back by future users, who would see the past user as a semi-transparent avatar interacting in the system.
Using this prototype, we designed an experiment to obtain insights on the following research question: 
\textit{Compared to synchronous and asynchronous temporal states, how does the bichronous temporal state impact the user experience of collaboration within a virtual environment?}

\subsection{Apparatus \& Technologies}
We opted to develop our system for use with the Meta Quest 3 head-worn display (HWD) and its touch controllers.
The HWD has a 108\textdegree\ horizontal and 98\textdegree\ vertical (43\textdegree\ upward and 55\textdegree\ downward) field of view and a resolution of $2064\times2208$ pixels per eye with a refresh rate of 90 Hz.
The native headset tracking allowed us to track users' movements in our system.
The Meta Quest 3 features integrated speakers and a microphone for audio input/output.

We created our system using version 2022.3.30f1 of Unity Game Engine.
We used version 2.13.0 of the Normcore networking library to enable multiple users to join the CVE, communicate with each other, and interact with virtual objects in real time.
We leveraged a locally-hosted server and the Secure File Transfer Protocol to access, download, and upload recordings.

\subsection{System}
Our system features a VE consisting of a ground plane within an empty skybox and a cuboid-lattice model positioned at its center, as shown in Fig. \ref{fig:teaser}A.
The model was generated using an open-source computed tomography dataset \cite{Klacansky_AMParts2022}.
Each user is represented in the VE by a low-poly avatar, as low-poly avatars support effective communication regardless of their abstract nature \cite{Roth_AvatarInteractionQuality2016}.
The avatar consists of the user's head, shown as a generic HWD, and their touch controllers, shown as generic oblong controllers.
Anchored above the avatar's HWD is a name label to identify the user (hidden in Fig. \ref{fig:teaser} for submission anonymity).
The avatar is colored to act as another identifiable trait.
An avatar's opacity distinguishes its temporal state: fully-opaque avatars are of real-time users and translucent avatars are recordings (see Fig. \ref{fig:teaser}C-E).

Users can physically move within an unobstructed 2 m $\times$ 2 m guardian boundary to perform smaller position and rotation adjustments.
Users can navigate within the VE using ray-cast teleportation from either of their controllers by pushing the controller's joystick forward to cast the ray, then releasing it to teleport to wherever the ray contacted the ground plane.
Multi-scale navigation is supported via the PRIMO technique \cite{Pavanatto_Primo2025}, in which users cast the ray at a specific octant of the cuboid-lattice model, causing them to be scaled down so they can view the selected octant more closely.
Pulling the joystick backward then releasing it when the user is at a smaller scale causes them to grow to their previous scale.

By pressing the primary button on either controller, users can toggle the view of the recording menu (Fig. \ref{fig:teaser}B).
This menu is populated with a list of recordings created by previous users, prefixed with the date of their creation.
Users may create recordings by selecting a ``Start Recording'' menu option fixed to the top of the recording menu using the trigger button on their controller.
This will close the recording menu and display a head-fixed recording icon (i.e., a red circle in the top-right corner of the viewport).
They can stop the recording and upload it for recording menu access by again opening the menu and selecting the top menu option, labeled as ``Stop Recording.''
Users can download a recording for playback by navigating in the menu using the joystick and pulling the trigger button to select the currently hovered menu option.
The menu is closed once the recording is downloaded successfully.
We chose to use avatar recording annotations due to their multi-modal nature, which provides benefits for collaborative tasks via speech and body gestures including deictic referencing and expressivity in CVEs \cite{Chow_ChallengesAsyncCollabVR2019}.

When a recording has been downloaded, pressing the primary button will toggle the playback control interface (Fig. \ref{fig:teaser}C).
The playback controls can be navigated by using the joystick and allow users to toggle the play/pause state of the recording, rewind by five seconds, fast-forward by five seconds, or close the recording.
When they select to close the recording, the playback control interface closes; the next time they press the primary button, the recording menu will display.
The CVE was symmetric between collaborators, such that a recording opened and interacted with by one collaborator was uniformly displayed to all joined users.
While it is possible to allow asymmetry of the CVE by providing unique instances for individualized viewership of recordings \cite{Lammert_StudyAnalyzer2024}, we wanted our CVE to resemble a physical space, where collaborators would be aware of all actions performed inside the CVE by their peers upon entering.

\subsection{Narrative \& Task} \label{ScenarioTask}
We created a study narrative in order to investigate the influence of temporal states on collaboration.
In this narrative, the participant was instructed to act as a quality assurance agent tasked with summarizing reported defects in a cuboid lattice model for a manufacturing company.
The task consisted of sequentially reviewing two previously recorded defect inspections in the VE.
To progress between recordings and to complete the task, the participant was required to verbally report the name and cause of the defect described and shown in the given recording. 
A live collaborator (confederate) joined in the middle of the first recording, which was made by themselves, causing a temporal state transition from Asynchronous to Bichronous.
After the participant reported the defect name and cause, they closed the recording and entered the Synchronous temporal state.
The live collaborator then helped the participant determine the second recording they needed to select, recorded by a different collaborator, thus causing a temporal state transition from Synchronous to Bichronous.
In the middle of the second recording, the live collaborator left the VE, causing a temporal state transition from Bichronous to Asynchronous.
Fig. \ref{fig:task} portrays the transitions between temporal states during the task.

\subsection{Design} \label{Design}
This exploratory study was within-subjects, where each participant completed the defect inspection task in two trials.
In the first trial, participants performed the task continuously, without any interruptions from the investigator.
In the second trial, the investigator interrupted the task at each temporal state transition to administer surveys (refer to Sec. \ref{Measures}).
The order in which participants viewed recordings was counterbalanced across trials.
The independent variables consisted of the \textit{Task Trial} (i.e., Continuous and Interrupted), \textit{Temporal State} (i.e., Asynchronous, Bichronous, and Synchronous), and \textit{Entity} (i.e., Annotation and Live Collaborator). 

\begin{figure}
    \centering
    \includegraphics[width=.95\linewidth]{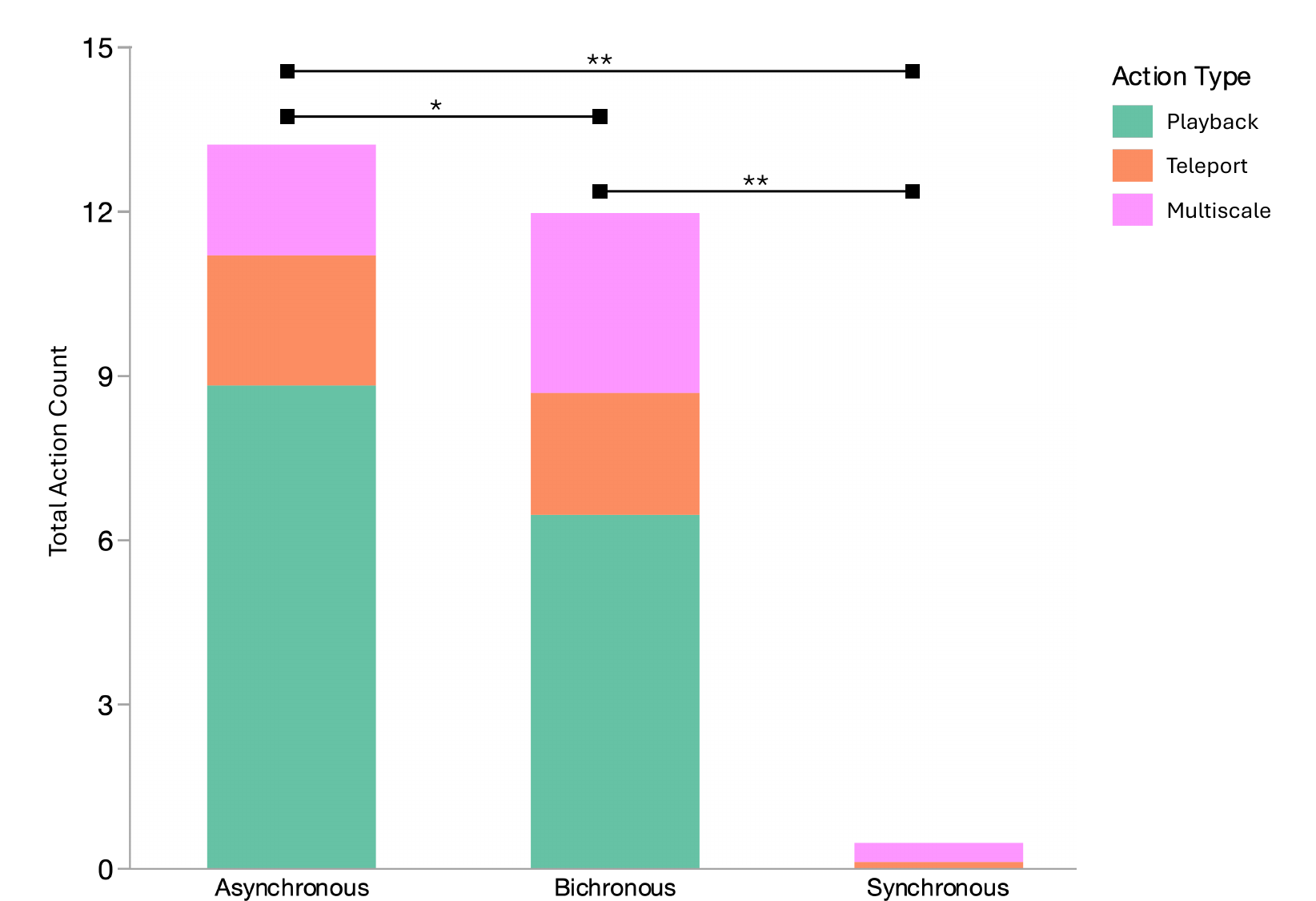}
    \caption{Mean control actions. Significantly different pairs are marked with * when $\emph{p} \le 0.05$ and ** when $\emph{p} \le 0.01$.}
    \Description{}
    \label{fig:actions}
\end{figure}

\subsection{Measures} \label{Measures}
We collected a variety of data regarding the social behaviors of participants per temporal state across objective and subjective measures.
As an individual's actions are dependent on their team members in a work group \cite{Andriessen_GroupwareBook2003}, we logged the number of playback, teleport, and multiscale navigation actions conducted during the study by the participants.
We additionally logged the position, orientation, and scale of the participant, annotation and live collaborator during the study, as variations in interpersonal space influence collaborative behaviors \cite{Hall_Proxemics1968, Williamson_DigitalProxemics2022}.
These were our objective measures.

Since attention has been identified as influential to the effectiveness of collaborative exchanges across temporal states \cite{Pidel_CollabSystematic2020}, we used three questions on a 7-point Likert scale to determine the participants' attention allocation during temporal states throughout the Interrupted Task Trial.
We additionally used three questions on a 7-point Likert scale to measure the participants' perceived behavioral interdependence to supplement the objective measure of total actions, subjectively.
Survey questions were administered separately for each collaborative entity. 
More specifically, participants completed one survey assessing attention allocation and behavioral interdependence after the Asynchronous temporal state regarding the annotation (see A.2), one after the Synchronous state regarding the live collaborator (see A.1), and two after the Bichronous state: one for the annotation and one for the live collaborator.
The attention allocation and behavioral interdependence questions used were each subsets taken from prior work for measuring these characteristics \cite{Harms_InternalSocialPresence2004}.

\begin{figure*}
        \centering
        \begin{subfigure}[b]{0.45\textwidth}  
            \centering 
            \includegraphics[width=\textwidth]{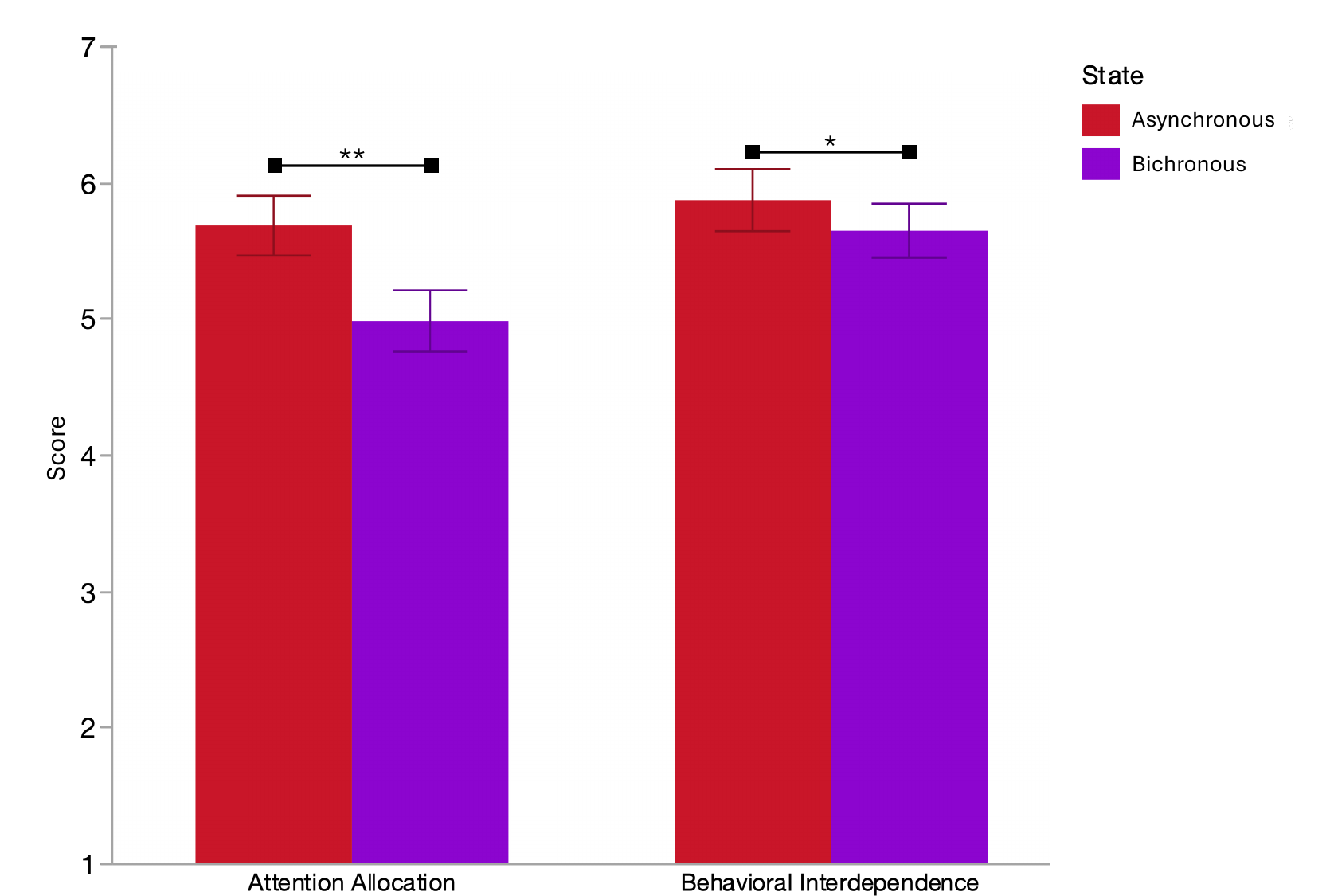}
            \caption[]%
            {{\small Annotation scores.}}
            \label{fig:ABIRec}
        \end{subfigure}
        \hfill
        \begin{subfigure}[b]{0.45\textwidth}   
            \centering 
            \includegraphics[width=\textwidth]{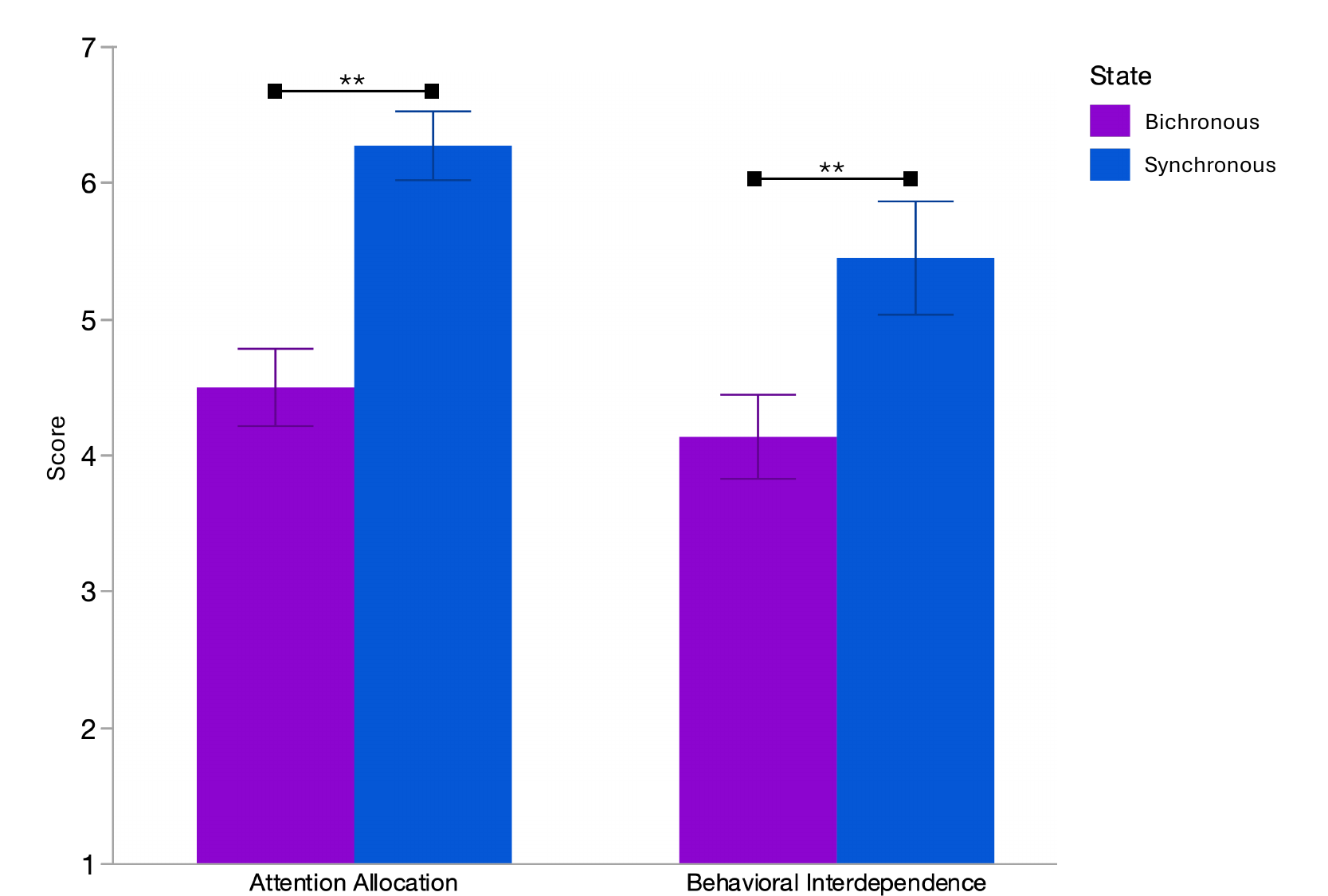}
            \caption[]%
            {{\small Live Collaborator scores.}}
            \label{fig:ABICon}
        \end{subfigure}
        \caption[]
        {The mean attention and behavioral interdependence scale scores per entity. Whiskers are the $\pm S.E.$ spread of the data.}
        \Description{}
        \label{fig:ABIs}
\end{figure*}

Finally, we recorded responses from a semi-structured interview conducted after completing the Continuous Task Trial (see A.3).
The questions posed were intended to provide context as to the impact of each temporal state on the collaborative experience for the participant.
These comprised our subjective measures.

\subsection{Participants}
Twenty participants were recruited for our study using email lists and newsletters.
All completed a screening questionnaire before being able to schedule a study session, stating they were: (1) 18 years of age or older, (2) had normal or corrected vision, (3) were able to walk and stand for long periods without assistance, and (4) were fluent in English.
The ages of the participants ranged from 20 to 36 years (M=23.95, SD=5.11). 
Ten participants were female and ten were male.
Five participants reported they had used VR once or twice, four had used VR three to ten times, and 11 had used VR more than ten times.
The study was approved by our local Institutional Review Board.

\subsection{Procedure}
The study followed a three-phase procedure: pre-study, study, and post-study.
In the pre-study phase, the participant was introduced to the investigator and the confederate, reviewed and signed a consent document, and was introduced to the equipment to be used during the study.
Their interpupillary distance was measured and set in the HWD, followed by adjustments to the HWD straps to fit the participant.
The participant was briefed on the study scenario and trained by the investigator on how to use the equipment, completing a series of actions where they used the HWD and controllers to interact with the system.
They then moved on to the study phase.

The study phase consisted of two task trials: the Continuous Task Trial and the Interrupted Task Trial.  
During the Continuous Task Trial, the participant followed the task requirements specified in Section \ref{ScenarioTask}.
However, no surveys were administered following collaborative state transitions.
Following completion of the Continuous Task Trial, the participant removed the HWD and responded to a semi-structured audio interview.
After the interview, the participant put the HWD back on and did the Interrupted Task Trial, where the scenario would pause after each collaborative state transition. 
During the pause, the participant verbally responded to attention allocation and behavioral interdependence Likert scale questions according to their most recent experienced collaborative state, resuming the review task after answering them.

Once the participant completed the Interrupted Task Trial, they moved on to the post-study phase, where they completed a background questionnaire on basic demographic information (see A.4). 
This concluded the post-study phase and the study.
The study procedure took approximately 60 minutes to complete.

\section{Results}

\begin{figure*}
        \centering
        \begin{subfigure}[b]{0.45\textwidth}
            \centering
            \includegraphics[width=\textwidth]{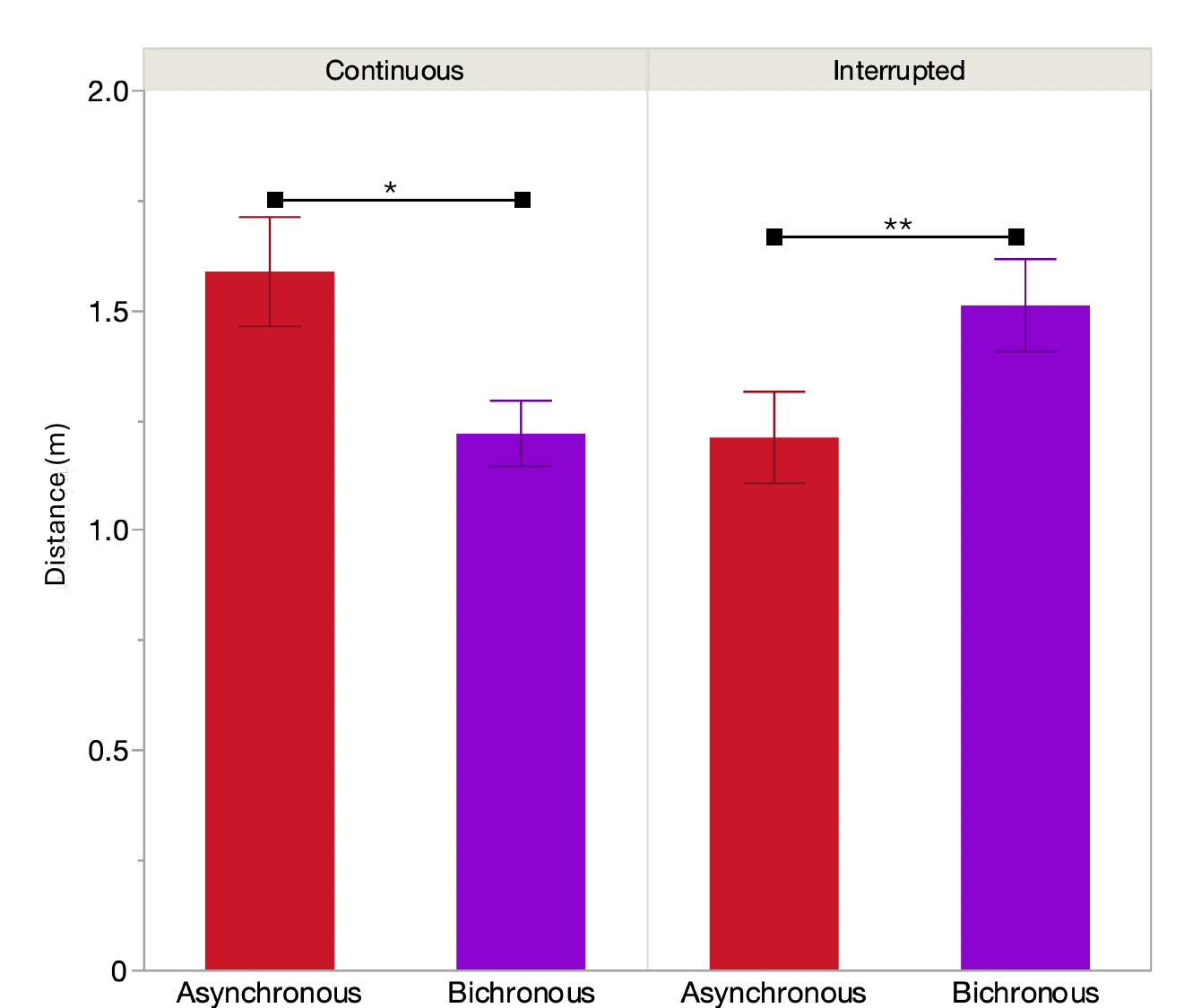}
            \caption[]%
            {{\small Distance between the participant and the annotation.}}
            \label{fig:ProxRec}
        \end{subfigure}
        \hfill
        \begin{subfigure}[b]{0.45\textwidth}   
            \centering 
            \includegraphics[width=\textwidth]{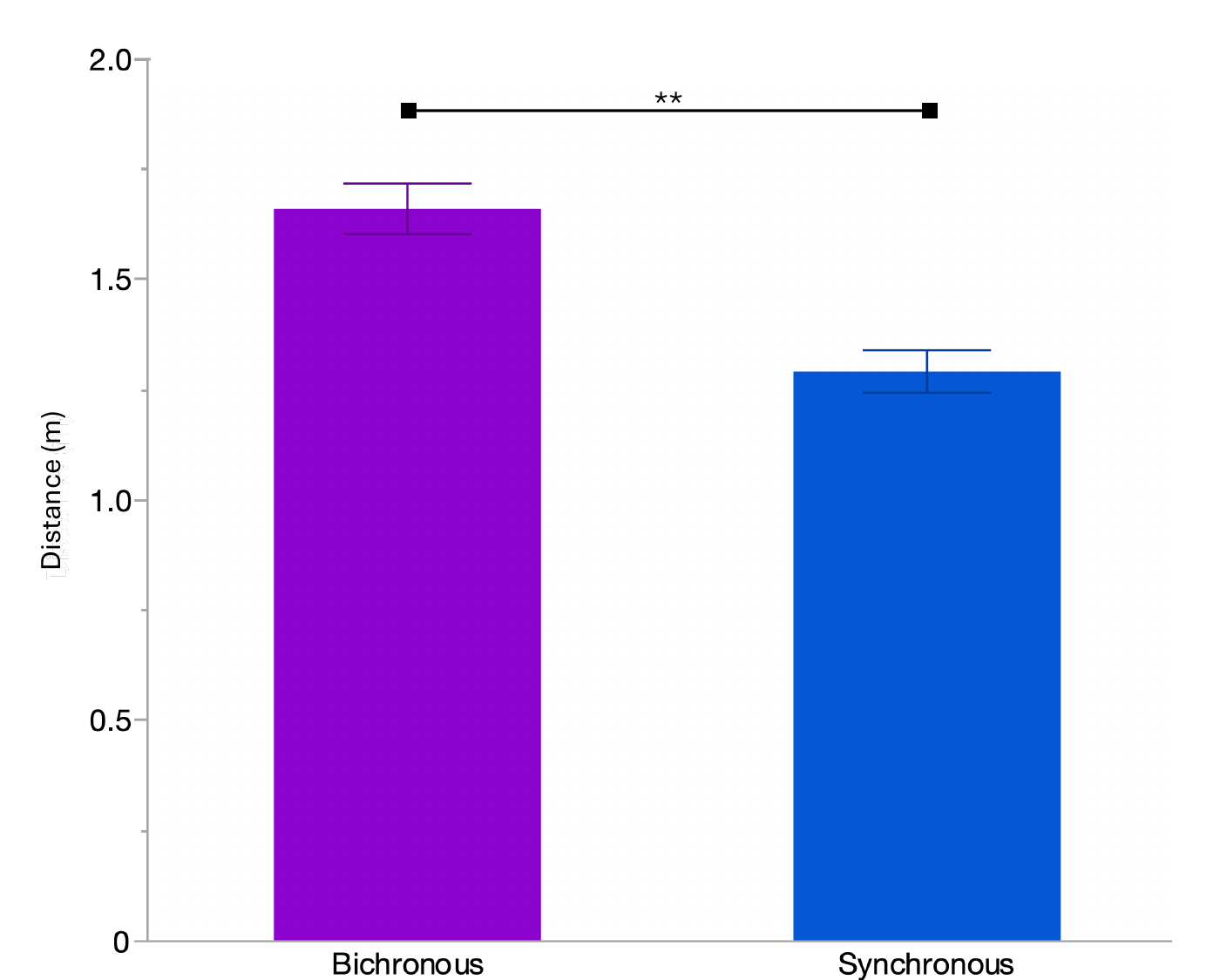}
            \caption[]%
            {{\small Distance between the participant and the live collaborator.}}  
            \label{fig:ProxCon}
        \end{subfigure}
        \caption[]
        {The mean distance of the participant to each entity per temporal state.}
        \Description{}
        \label{fig:ProxAll}
\end{figure*}

\subsection{Control Action Count} \label{Results:Action}
As the sample size of the control action count exceeded 50, a Kolmogorov-Smirnoff goodness-of-fit test was performed to test normality for each entity \cite{Mishra_DescriptiveStats2019}.
The data was not normally distributed, so we applied an Aligned Rank Transform (ART) \cite{Wobbrock_ArtAnova2011} before using an Analysis of Variance (ANOVA) for both.
If any statistically significant effect was reported by the ART-ANOVA, we performed post-hoc pairwise comparisons using the ART-C procedure with False Discovery Rate (FDR) corrections \cite{Elkin_ArtC2021}.

The ART-ANOVA indicated a significant main effect of the Temporal State ($F(2, 175) = 70.79$, $\emph{p} < 0.01$, $\eta^2p = 0.45$) and no effect of the Trial ($F(1, 175) = 0.52$, $\emph{p} = 0.47$) or the interaction of the Trial and the Temporal State ($F(2, 175) = 0.01$, $\emph{p} = 0.99$).
Post-hoc comparisons for the Temporal State indicated significant differences for all pairs.
Participants performed significantly fewer actions during the Synchronous temporal state compared to both the Asynchronous ($\emph{p} < 0.01$) and Bichronous ($\emph{p} < 0.01$) temporal states. 
They also performed fewer actions during the Bichronous state compared to the Asynchronous state ($\emph{p} = 0.04$).
The mean count per action type and the significant pairings are shown in Fig. \ref{fig:actions}.

We conducted additional analyses for each type of action.
The ART-ANOVA for the Playback action indicated a significant main effect of the Temporal State ($F(2, 175) = 94.99$, $\emph{p} < 0.01$, $\eta^2p = 0.52$) and no effect of the Trial ($F(1, 175) = 2.54$, $\emph{p} = 0.11$) or the interaction of the Trial and the Temporal State ($F(2, 175) = 0.89$, $\emph{p} = 0.41$).
Post-hoc comparisons for the Temporal State indicated that participants performed significantly fewer playback actions during the Synchronous temporal state compared to both the Asynchronous ($\emph{p} < 0.01$) and Bichronous ($\emph{p} < 0.01$) temporal states.

The ART-ANOVA for the Multiscale action indicated a significant main effect of the Temporal State ($F(2, 175) = 36.15$, $\emph{p} < 0.01$, $\eta^2p = 0.29$) and the Trial ($F(1, 175) = 3.99$, $\emph{p} = 0.05$, $\eta^2p = 0.02$), but no effect for the interaction of the Trial and the Temporal State ($F(2, 175) = 0.87$, $\emph{p} = 0.42$).
Given the small effect of the Trial, we only conducted post-hoc comparisons for the Temporal State.
Participants performed significantly fewer multiscale actions during the Synchronous temporal state compared to both the Asynchronous ($\emph{p} < 0.01$) and Bichronous ($\emph{p} < 0.01$) temporal states.
They also performed fewer multiscale actions during the Asynchronous state compared to the Bichronous state ($\emph{p} < 0.01$).

The ART-ANOVA for the Navigation action indicated a significant main effect of the Temporal State ($F(2, 175) = 32.52$, $\emph{p} < 0.01$, $\eta^2p = 0.27$) and the interaction of the Trial and the Temporal State ($F(2, 175) = 3.03$, $\emph{p} = 0.05$, $\eta^2p = 0.03$), but no effect of the Trial ($F(1, 175) = 1.16$, $\emph{p} = 0.28$).
Given the small effect of the interaction between Trial and Temporal State, we only conducted post-hoc comparisons for the Temporal State.
Participants performed significantly fewer navigation actions during the Synchronous temporal state compared to both the Asynchronous ($\emph{p} < 0.01$) and Bichronous ($\emph{p} < 0.01$) temporal states.

\subsection{Attention \& Behavioral Interdependence} \label{Results:ABI}
We analyzed the Temporal States for each entity for our attention and perceived behavioral interdependence scales.
We conducted a non-parametric analysis on the Likert scale responses using the Wilcoxon signed-rank test for each of the scales.
The mean scores per scale and the significant pairings are shown in Fig. \ref{fig:ABIs}.

\paragraph{Annotation} Statistically significant differences emerged between temporal states, with the Bichronous state showing lower attention allocation ($\emph{Z} = 3.00$, $\emph{p} < 0.01$, $r_{rb} = 0.69$) and behavioral interdependence scores ($\emph{Z} = 2.26$, $\emph{p} = 0.02$, $r_{rb} = 0.58$) compared to the Asynchronous state.

\paragraph{Live Collaborator} Statistically significant differences emerged between temporal states, with the Bichronous state showing lower attention allocation ($\emph{Z} = -3.31$, $\emph{p} < 0.01$, $r_{rb} = -0.76$) and behavioral interdependence scores ($\emph{Z} = -3.43$, $\emph{p} < 0.01$, $r_{rb} = -0.83$) compared to the Synchronous state.

\subsection{Proxemics} \label{Results:Prox}
The sample sizes of the mean distances between the participant and the other entities exceeded 50, so a Kolmogorov-Smirnoff goodness of fit test was performed across the Continuous and Interrupted Task Trials and Temporal States to test normality.
The data was not normally distributed, so we performed an ART-ANOVA.
If any statistically significant effect was reported by the ART-ANOVA, we performed post-hoc pairwise comparisons using the ART-C procedure with FDR corrections.
The mean distances and the significant pairings are shown in Fig \ref{fig:ProxAll}.

\paragraph{Annotation} Looking first at the distances between the participant and the annotation, the ART-ANOVA indicated a significant interaction effect between the Trial and the Temporal State ($F(1, 137) = 17.65$, $\emph{p} < 0.01$, $\eta^2p = 0.11$) and no main effects of the Trial ($F(1, 137) = 0.89$, $\emph{p} = 0.34$) or the Temporal State ($F(1, 137) = 0.31$, $ \emph{p} = 0.58$).
Post-hoc comparisons for the Trial $\times$ Temporal State interaction revealed significantly greater distance between the participant and the annotation in the $Asynchronous \times Continuous$ condition compared to the $Bichronous \times Continuous$ condition ($\emph{p} = 0.03$). 
In contrast, the $Bichronous \times Interrupted$ condition showed significantly greater distance than the $Asynchronous \times Interrupted$ condition ($\emph{p} < 0.01$).

\paragraph{Live Collaborator} Considering the distance between the participant and the live collaborator, the ART-ANOVA reported a significant main effect of the Temporal State ($F(1, 97) = 26.70$, $\emph{p} < 0.01$, $\eta^2p = 0.22$) and no effect of the Trial ($F(1, 97) = 3.28$, $\emph{p} = 0.07$) or the interaction of the Trial and the Temporal State ($F(1, 97) = 0.95$, $\emph{p} = 0.33$).
Post-hoc comparisons for the Temporal State revealed significantly greater distance between the participant and the live collaborator in the Bichronous state compared to the Synchronous state ($\emph{p} < 0.01$).

\subsection{Qualitative Analysis}
We conducted an inductive thematic analysis of the 20 semi-structured interviews.
The interview audio recordings were first transcribed in Microsoft Word Online.
This was followed by an initial reading of the transcripts to identify recurring phrases and concepts, which were given general descriptors as themes.
These themes were than organized in spreadsheets using Microsoft Excel Online. 
The themes were iteratively refined, with their underlying content being assigned to specific sub-themes.
We briefly describe the themes and sub-themes in this subsection.

\subsubsection{Interactional Constraints}
This theme arose from participants reporting changed interaction behavior due to collaborator crowding and annotation clutter per temporal state.
It encompasses two sub-themes: social pressures and visual occlusion.

\paragraph{Social pressures}
13 participants described a difference in the way they behaved while observing the annotation in the presence of a live collaborator during the Bichronous temporal state.
\textit{P2} explained ``if there's one part that I'm trying to get more detail on and everyone else is caught up, then I wouldn't want to slow it all down,'' \textit{P11} described ``when the live collaborator is there ... I could potentially be wasting someone else's time,'' and \textit{P14} stated ``you feel pressured because someone is there waiting for you.''
Similarly for the Synchronous temporal state, \textit{P6} stated that just the live collaborator made them ``feel a little bit of pressure at times'' and \textit{P10} said ``the presence of [the live collaborator] in the long term would be bothering.''
However, when in the Asynchronous temporal state, participants indicated that they were more comfortable performing playback control actions and felt they had more time.
\textit{P6} reported ``I can rewind as many times and, I guess on a personal level, I don't feel as judged for rewinding so many times'' and \textit{P8} mentioned ``I felt when there is no live collaborator, ... I felt more free to go forward and back in the playback.''

\paragraph{Visual occlusion}
While performing collaborative activities in the Bichronous temporal state, 11 participants indicated that visual obstructions were an issue.
\textit{P1} stated when ``I was trying to follow the recording and my collaborator was standing next to me, I felt that I may collide with them or obstruct their view'' and \textit{P15} explained ``it was kind of hard to view the defect and be in a good spot to see everything [that] was going on.''
A similar concern of visual obstruction when working with only the live collaborator was brought up for the Synchronous temporal state by \textit{P8}: ``if you're the same height [as the live collaborator], your headset can get in the way.''
However, when describing the annotation representation during the Asynchronous temporal state without a live collaborator, participants reported no such issues. 
\textit{P5} stated ``I could visually see it and I could follow where to go along with the recording'' and \textit{P14} described that ``the recording was pretty nice because you were able to see them in the VR environment.''

\subsubsection{Information processing}
This theme developed based on participant feedback that information recall and understanding during collaboration were impacted by the temporal state.
It has two sub-themes: information recall and information understanding.

\paragraph{Information recall}
When reviewing the annotations, 15 participants noted the difficulty of locating information in an efficient manner during the Asynchronous temporal state to complete the task.
\textit{P13} described that they would ``lose track of some information that I need'' and \textit{P17} remarked that they would ``just go through [the annotation] a couple times to get the information back and also just to remember the answer.''
Conversely, the Bichronous temporal state enabled participants to find information more reliably with the help of their live collaborator.
\textit{P12} noted that it was ``nice to have the live collaborator there so that if I had a question about the [annotation], they can direct me to what the [annotation] is talking about'' and \textit{P20} stated that if they needed ``to find the part that talked about specific information, [the live collaborator] saves you more time.''
In this regard, the Bichronous temporal state is akin to the Synchronous temporal state, where conversations with the live collaborator could help steer the participant toward required information.
When discussing the Synchronous temporal state, \textit{P5} stated having the live collaborator gave them ``immediate feedback, so you're able to go and get the information that you need immediately'' and \textit{P12} mentioned the live collaborator could ``walk me through showing me what I'm trying to look for.''

\paragraph{Information understanding}
While annotations provide an understanding of an asynchronous contribution, they are static units of information that cannot provide clarification beyond their contents.
17 participants mentioned this issue throughout the semi-structured interview, with \textit{P4} stating if ``you don't understand a certain part of [the annotation] and you review it a couple of times and you still don't understand ... you're basically stuck'' and \textit{P9} remarking ``all the information that's there is all that will ever be there.''
In contrast, the Bichronous temporal state provides an advantage to the Asynchronous temporal state, as the annotation provides a baseline knowledge to guide discussion between collaborators, divulging new context and understanding in real time.
Participants described this, with \textit{P5} saying in the bichronous they could review the annotation ``with [the live collaborator] to explain context or what's going on'' and \textit{P13} stating they could review ``the detailed description of the whatever you are looking at and also you can ask questions if you don't understand.''
Furthermore, with the absence of the annotation during the Synchronous temporal state, no knowledge baseline is established to guide initial discussion as in the Bichronous temporal state.
\textit{P16} noted that they would not know what to talk about ``if the collaborator never talked about the [defect] causes'' and \textit{P17} said ``the collaborator won't know what you want unless you know what to ask.''

\section{Discussion} \label{Discussion}
In this section, we identify distinguishing characteristics of the Bichronous temporal state found by our user study.
We initially base these characteristics from our qualitative analysis themes, using the quantitative data and CSCW research to support them.

\subsection{Bichronous Amplifies Interactional Constraints}
Participants reported two distinct drawbacks when collaborators and annotations populated the CVE during the user study task in the Bichronous temporal state.
The first drawback of bichronous collaboration was the social pressure participants felt when reviewing annotations with a live collaborator. 
While assistance from the collaborator may have reduced the need for playback actions, interviews revealed that participants felt self-conscious, as if they were taking time away from the collaborator.
As a result, they engaged in significantly fewer playback actions during the Bichronous temporal state, as compared to the asynchronous state when no live collaborator was present. 
Participants in the Bichronous temporal state showed a significantly lower Total Action Count compared to the Asynchronous temporal state, reinforcing their claimed change in interactive behavior. 
The presence of a real-time observer appeared to alter participants’ interaction patterns in the CVE, similar to the Hawthorne effect during synchronous collaboration \cite{Franke_Hawthorne1978, Mayo_Hawthorne2004}. 
Although such pressure might increase task efficiency \cite{Erickson_SociallyTranslucentSystems1999}, it can negatively impact user experience by making participants feel restricted in how they interact with the system.

The second drawback of bichronous collaboration was increased visual occlusion during the study task.
Participants explained during the semi-structured interviews that the simultaneous presence of a live collaborator and annotation in the CVE disrupted their positioning, as they consciously tried to avoid colliding with either entity.
This impacted their ability to maintain an optimal view inside the CVE. 
This reported impact is supported by Proxemics data, which show participants keeping greater distance from both the annotation and live collaborator during the Bichronous temporal state compared to the other temporal states. 
While visual occlusion is also a known issue in synchronous and asynchronous collaboration \cite{Fraser_RealitiesCollabVR2000, Chen_MultiCollabVE2015, Chow_ChallengesAsyncCollabVR2019}, the bichronous temporal state appears to exacerbate it by combining both sources of interference within the same space.

\subsection{Bichronous Improves Information Processing}
Participants outlined two beneficial impacts on information processing while in the Bichronous temporal state.
The first benefit of bichronous collaboration was participants' improved ability to recall information during the study task. 
In semi-structured interviews, participants noted that the live collaborator helped direct them to details embedded in the asynchronous annotation that they may have otherwise forgotten. 
This enhancement is reflected in their perceived Behavioral Interdependence: rather than relying solely on one collaborator, participants in the Bichronous temporal state drew on both live and annotated contributions to locate relevant information.
This strategy aligns with prior work regarding synchronous information search, where collaborators maintain mutual awareness but often carry out the search independently \cite{Foley_SyncInfoRetrieval2010}.
Bichronous collaboration scaffolds this dynamic more effectively by supporting both independent exploration using the annotation and real-time prompting of the live collaborator.

The second benefit of bichronous collaboration was participants’ improved ability to understand the information presented during the study task. 
Participants explained that while the annotation in the Asynchronous temporal state helped them ground their knowledge, it lacked the ability to expand or clarify its content. 
In contrast, the synchronous state offered real-time interaction, but participants often felt uncertain about how to initiate discussion without a shared starting point.
This issue was resolved in the Bichronous temporal state: participants used the annotation to establish a baseline understanding, and then turned to the live collaborator to ask clarifying questions and explore the material in greater depth. 
This finding is supported by results from the Attention Allocation survey, which showed that the Bichronous temporal state required significantly less attention for managing both the annotation and the live collaborator compared to the Asynchronous and Synchronous temporal states. 
Participants were able to efficiently divide their attention by first grounding themselves in the annotation content and then using the live collaborator to enrich their comprehension.
Prior work regarding collaborative literature review supports this perceived improvement in information comprehension \cite{Rahman_CSR2015}, suggesting that performance benefits could also be garnered from bichronous collaboration.

\section{Limitations \& Future Work}
Our work had various limitations.
First, our study task did not fully assess the Synchronous temporal state comparably to the Asynchronous or Bichronous temporal states.
Participants only discussed with the live collaborator what annotation to review next, rather than being guided by the live collaborator through a defect's name and cause.
Future work should have the Synchronous temporal state task match the other temporal states.
Second, a confederate acted as the live collaborator for the study.
Future work should examine two or more participants working in the bichronous temporal state together to complete a shared task with equal expertise.
This would help to examine a peer-to-peer relationship, where the participants learn and discuss the information inside the CVE together.
Third, the study task was fixed (i.e., find the defect name and cause in an annotation).
Future work should investigate different tasks to examine how difficulty influences user behaviors across the temporal states.
Finally, the system had participants review only avatar annotations.
Annotation media such as text, speech, or images could be used to relay contributions instead.
Differences in visual occlusion and social pressures may exist between the media types.
Future work should investigate the use of other annotations and their influence on the temporal states.

\section{Conclusion}
We revisited the time-space matrix of CSCW and proposed the time continuum, including the bichronous state.
Our exploratory evaluation indicated that the bichronous temporal state is beneficial to collaborative activities for information processing, but has drawbacks such as changed interaction and positioning behaviors in the CVE.

\begin{acks}
This work was performed under the auspices of the U.S. Department of Energy by Lawrence Livermore National Laboratory under Contract DE-AC52-07NA27344.
\end{acks}

\balance
\bibliographystyle{ACM-Reference-Format}
\bibliography{_main}

\appendix
\onecolumn

\section{Questionnaires \& Interviews}

\subsection{Confederate Attention Allocation and Behavioral Interdependence Survey} \label{App:ConfSurvey}
    \hspace{10px}Participant ID
    \begin{tabular}{@{}p{.5in}p{4in}@{}}
        & \hrulefill \\
    \end{tabular}

    Transition State
    \begin{enumerate}
        \item [\cbox{}] Asynchronous
        \item [\cbox{}] Bichronous
        \item [\cbox{}] Synchronous
    \end{enumerate}

\begin{table}[H]
\begin{tabular}{lccccccc}
 & \begin{tabular}[c]{@{}c@{}}Strongly \\ Disagree\end{tabular} &  &  &  &  &  & \begin{tabular}[c]{@{}c@{}}Strongly \\ Agree\end{tabular} \\
I remained focused on the live collaborator throughout our interaction.       & \cbox{} & \cbox{} & \cbox{} & \cbox{} & \cbox{} & \cbox{} & \cbox{} \\
My behavior was often in direct response to the live collaborator's behavior. & \cbox{} & \cbox{} & \cbox{} & \cbox{} & \cbox{} & \cbox{} & \cbox{} \\
The live collaborator's thoughts were clear to me.                            & \cbox{} & \cbox{} & \cbox{} & \cbox{} & \cbox{} & \cbox{} & \cbox{} \\
The  live collaborator did not receive my full attention.                     & \cbox{} & \cbox{} & \cbox{} & \cbox{} & \cbox{} & \cbox{} & \cbox{} \\
Understanding the  live collaborator was difficult.                           & \cbox{} & \cbox{} & \cbox{} & \cbox{} & \cbox{} & \cbox{} & \cbox{} \\
My behavior was closely tied to the  live collaborator's behavior.            & \cbox{} & \cbox{} & \cbox{} & \cbox{} & \cbox{} & \cbox{} & \cbox{}
\end{tabular}
\end{table}

\subsection{Annotation Attention Allocation and Behavioral Interdependence Survey Survey} \label{App:AnnotSurvey}
    \hspace{10px}Participant ID 
    \begin{tabular}{@{}p{.5in}p{4in}@{}}
        & \hrulefill \\
    \end{tabular}

    Transition State
    \begin{enumerate}
        \item [\cbox{}] Asynchronous
        \item [\cbox{}] Bichronous
        \item [\cbox{}] Synchronous
    \end{enumerate}

\begin{table}[H]
\begin{tabular}{lccccccc}
 &
  \begin{tabular}[c]{@{}c@{}}Strongly \\ Disagree\end{tabular} &
   &
   &
   &
   &
   &
  \begin{tabular}[c]{@{}c@{}}Strongly \\ Agree\end{tabular} \\
I remained focused on the recorded avatar throughout our interaction. &
  \cbox{} &
  \cbox{} &
  \cbox{} &
  \cbox{} &
  \cbox{} &
  \cbox{} &
  \cbox{} \\
My behavior was often in direct response to the recorded avatar's behavior. &
  \cbox{} &
  \cbox{} &
  \cbox{} &
  \cbox{} &
  \cbox{} &
  \cbox{} &
  \cbox{} \\
The recorded avatar's thoughts were clear to me.       & \cbox{} & \cbox{} & \cbox{} & \cbox{} & \cbox{} & \cbox{} & \cbox{} \\
The recorded avatar did not receive my full attention. & \cbox{} & \cbox{} & \cbox{} & \cbox{} & \cbox{} & \cbox{} & \cbox{} \\
Understanding the recorded avatar was difficult.       & \cbox{} & \cbox{} & \cbox{} & \cbox{} & \cbox{} & \cbox{} & \cbox{} \\
My behavior was closely tied to the recorded avatar's behavior. &
  \cbox{} &
  \cbox{} &
  \cbox{} &
  \cbox{} &
  \cbox{} &
  \cbox{} &
  \cbox{}
\end{tabular}
\end{table}

\subsection{Semi-Structured Interview} \label{App:Int}
\begin{enumerate}
    \item Did you change how you interacted with the playback controls when alone versus with a collaborator?
    \item How did you feel about watching the recording with the person who created it?
    \item How did you feel when you watched a recording with somebody else?
    \item What are the pros and cons of working with a...
    \begin{enumerate}
        \item Live collaborator?
        \item Recording?
        \item Both a live collaborator and recording?
    \end{enumerate}
\end{enumerate}

\pagebreak
\subsection{Background Questionnaire} \label{App:Back}
    \hspace{10px}Participant ID 
    \begin{tabular}{@{}p{.5in}p{4in}@{}}
        & \hrulefill \\
    \end{tabular}
    
    Sex
    \begin{enumerate}
        \item [\cbox{}] Male
        \item [\cbox{}] Female
        \item [\cbox{}] Prefer not to say
    \end{enumerate}

    Age
    \begin{tabular}{@{}p{.5in}p{4in}@{}}
        & \hrulefill \\
    \end{tabular}

    What is your profession?
    \begin{tabular}{@{}p{.5in}p{4in}@{}}
        & \hrulefill \\
    \end{tabular}

    Do you wear glasses or contact lenses?
    \begin{enumerate}
        \item [\cbox{}] Neither
        \item [\cbox{}] Glasses
        \item [\cbox{}] Contact Lenses
    \end{enumerate}

    Are you:
    \begin{enumerate}
        \item [\cbox{}] Right-handed
        \item [\cbox{}] Left-handed
        \item [\cbox{}] Ambidextrous
    \end{enumerate}

\begin{table}[H]
\begin{tabular}{lccccccc}
 &
  \begin{tabular}[c]{@{}c@{}}Very\\ Tired\end{tabular} &
   &
   &
   &
   &
   &
  \begin{tabular}[c]{@{}c@{}}Not\\ Tired\end{tabular} \\
Rate your fatigue level: &
  \cbox{} &
  \cbox{} &
  \cbox{} &
  \cbox{} &
  \cbox{} &
  \cbox{} &
  \cbox{}
\end{tabular}
\end{table}

    How many times have you tried virtual or augmented reality?
    \begin{enumerate}
        \item [\cbox{}] Never used
        \item [\cbox{}] Once or twice
        \item [\cbox{}] 3 to 10 times
        \item [\cbox{}] More than 10 times
    \end{enumerate}

\begin{table}[H]
\begin{tabular}{lccccccc}
 &
  \begin{tabular}[c]{@{}c@{}}Not at\\ All\end{tabular} &
   &
   &
   &
   &
   &
  \begin{tabular}[c]{@{}c@{}}Very\\ Often\end{tabular} \\
How frequently do you collaborate with others remotely? &
  \cbox{} &
  \cbox{} &
  \cbox{} &
  \cbox{} &
  \cbox{} &
  \cbox{} &
  \cbox{}
\end{tabular}
\end{table}

\end{document}